%% file: main.tex
\newif\ifusenix
\newif\ifacm
\newif\ifmcom
\newif\ifimwut

\usenixfalse
\acmtrue
\mcomfalse
\imwutfalse

\input{preamble}
\begin{document}
\title{Densify \& Conquer: Densified, smaller base-stations can conquer the increasing carbon footprint problem in nextG wireless}
\renewcommand{\shorttitle}{DensQuer}
\newcommand{\name}{DensQuer\xspace}
\author{Agrim Gupta, Adel Heidari, Jiaming Jin, Dinesh Bharadia\\
\{agg003,adheidari,j8jin,dineshb\}@ucsd.edu\\
University of California, San Diego}

\input{0-abstract}
\maketitle



\input{1-intro_ue_v1}

\input{2-background_ue_v1}
\input{3-design_ue_v2}
\input{5-evaluation.tex}
\input{6-limitations.tex}
\input{7-related.tex}
\input{8-conclusion.tex}
\input{9-acknowledge}
\bibliographystyle{unsrt}
\bibliography{main}

\end{document}

%% file: preamble.tex
\ifusenix
	\documentclass[letterpaper,twocolumn,9pt]{article}
	\usepackage{usenix}
	\usepackage{times}
\fi

\ifacm
\documentclass[sigconf,9pt]{acmart}

\usepackage[english]{babel}
\usepackage{blindtext}
\usepackage{times}

\renewcommand\footnotetextcopyrightpermission[1]{} 
\setcopyright{none}

\acmDOI{}

\acmISBN{}

\renewcommand\footnotetextcopyrightpermission[1]{}

\acmConference[ArXiv preprint]{ArXiv Preprint}{}

\acmPrice{}
\fi

\ifmcom
  \documentclass[sigconf]{sig-alternate-9pt}
\fi
\usepackage{xcolor}
\usepackage{amsfonts}
\PassOptionsToPackage{hyphens}{url}\usepackage{hyperref}
\usepackage{color}
\usepackage{graphics}
\usepackage{graphicx}
\usepackage{url}
\usepackage{listings}
\usepackage{multicol}
\usepackage{multirow}
\usepackage[scaled]{helvet}
\usepackage{rotating}
\usepackage{xspace}
\usepackage[ruled,vlined]{algorithm2e}
\usepackage{comment}
\usepackage{enumitem}
\usepackage{amsmath}
\usepackage{mathrsfs}
\usepackage{cancel}
\usepackage{cleveref}
\usepackage{subcaption}
\usepackage{float}
\usepackage[font=small,labelfont=bf]{caption}
\usepackage[skip=-1pt]{subcaption}
\usepackage[explicit,noindentafter]{titlesec}
\titlespacing\section{2pt}{4pt plus 2pt minus 2pt}{4pt plus 2pt minus 2pt}
\titlespacing\subsection{2pt}{4pt plus 2pt minus 2pt}{4pt plus 2pt minus 2pt}
\titlespacing\subsubsection{2pt}{4pt plus 2pt minus 2pt}{4pt plus 2pt minus 2pt}

%% file: 0-abstract.tex
\begin{abstract}

\noindent 
Connectivity on-the-go has been one of the most impressive technological achievements in the 2010s decade.
However, multiple studies show that this has come at an expense of increased carbon footprint, that also rivals the entire aviation sector's carbon footprint.
The two major contributors of this increased footprint are (a) smartphone batteries which affect the embodied footprint and (b) base-stations that occupy ever-increasing energy footprint to provide the last mile wireless connectivity to smartphones.
The root-cause of both these turn out to be the same, which is communicating over the last-mile lossy wireless medium.
We show in this paper, titled \name, how base-station densification, which is to replace a single larger base-station with multiple smaller ones, reduces the effect of the last-mile wireless, and in effect conquers both these adverse sources of increased carbon footprint.
Backed by a open-source ray-tracing computation framework (Sionna), we show how a strategic densification strategy can minimize the number of required smaller base-stations to practically achievable numbers, which lead to about 3x power-savings in the base-station network.
Also, \name is able to also reduce the required deployment height of base-stations to as low as 15m, that makes the smaller cells easily deployable on trees/street poles instead of requiring a dedicated tower.
Further, by utilizing newly introduced hardware power rails in Google Pixel 7a and above phones, we also show that this strategic densified network leads to reduction in mobile transmit power by 10-15 dB, leading to about 3x reduction in total cellular power consumption, and about 50\% increase in smartphone battery life when it communicates data via the cellular network.

\end{abstract}

%% file: 1-intro_ue_v1.tex
\section{Introduction}\label{sec:intro}

Over the past-decade, cellular networks have enabled almost pervasive connectivity, enabling multitiude of applications like video call to social media to gaming enjoyed all on a platform.
Further, people expect to be connected to the network 24/7, and do not want to miss the latest messages/e-mails/news as they are on the go.
Two primary enablers of these are smartphone mobiles, and the base-station mobile is connected to, which facilitates a wireless connection of mobile device to the rest of the world.
As we moved from 4G to 5G, there are varied reports on mobile devices experiencing reduced battery life~\cite{han2020energy,xu2020understanding}, and as well an increase in power bills faced by network companies for base-station operation~\cite{bts_ericsson,bts70,bts80}.
A larger overarching problem is the worsening carbon footprint of the larger telecom and information network, which now rivals that of the aviation sector~\cite{footprint2}.
Unsurprisingly, the two largest contributors are the embodied carbon footprint in smartphones, and the operational carbon footprint in base-station network~\cite{footprint1}.
Hence, it is the need of the hour while planning for 6G and beyond, to minimize the impact of cellular networks on the smartphone battery and base-station power consumption, the highest contributors of embodied and operational footprints respectively~\cite{mck,consumer1}.

Enabling wireless last-mile connectivity forms a major part of the energy consumption at both the smart-phone and the base-stations.
This is because instead of a controlled wired-medium, wireless connectivity is over the air, which is a lossy medium. The physics behind this phenomenon lies in the fact that the signal is radiated in all directions from the source, leading to a natural dispersion of the electromagnetic waves. As the signal propagates through space, it encounters various obstacles and undergoes absorption, reflection, and scattering. This phenomenon creates substantial signal losses as the base-station and smartphones communicate wirelessly. 
This translates to high mobile battery drainage at the phone, and increased power-draw from the mains at the base-station, to transmit large-enough power data packets to travel through the air and get received successfully at a far-off distance.
Fundamentally, to enable greener energy footprint of base-stations, and positively increase the operation time on a single battery charge for smartphones in cellular network,
these large wireless signal losses through the air need to be minimized.

\begin{figure}
    \centering
    \includegraphics[width=0.5\textwidth]{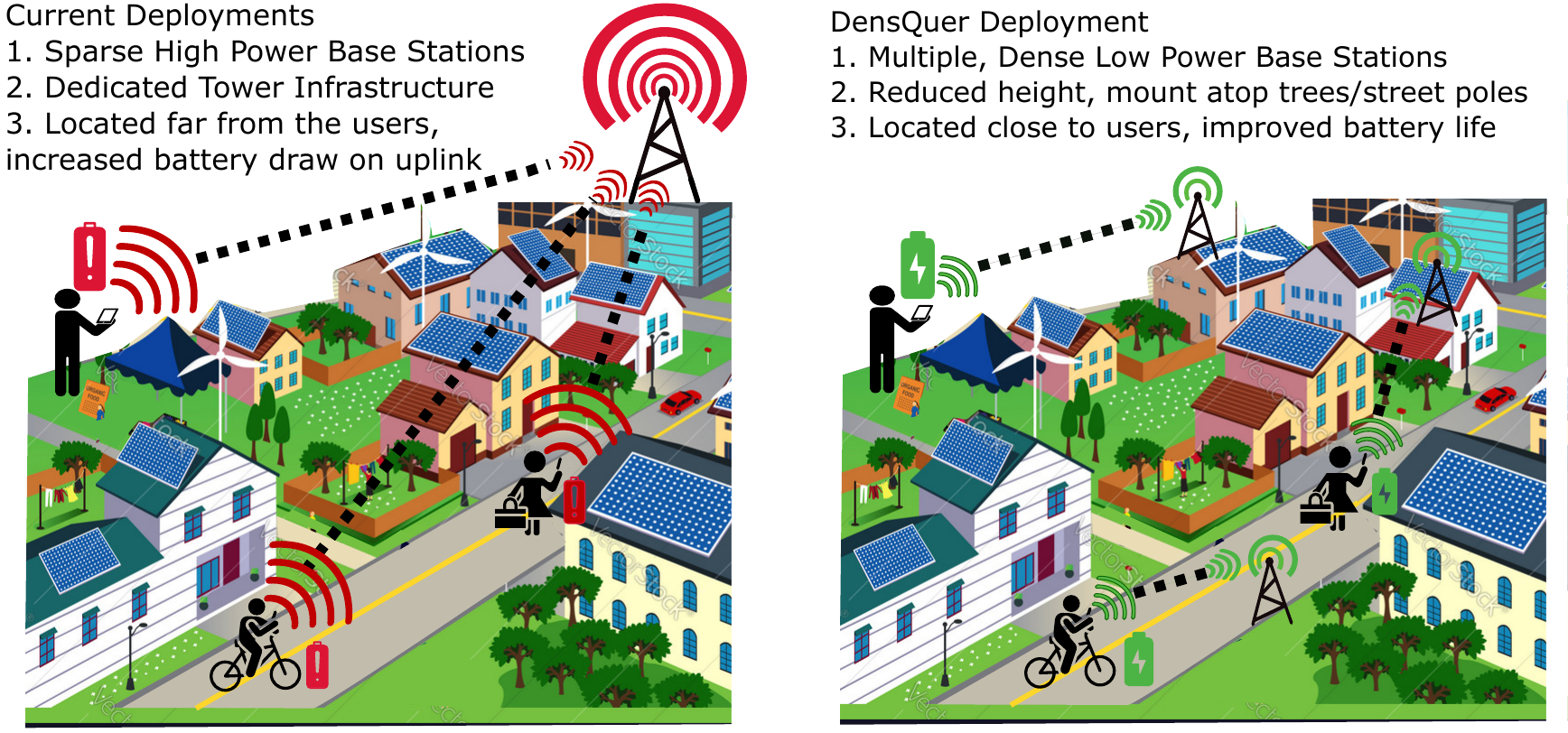}
    \caption{In an urban cellular coverage, a single macro base-station has to overcome large wireless path loss, which can equivalently be served with smaller multiple low-power base-stations, that can be located closer at lower heights, which has positive effect on smartphone battery as well}
    \label{fig:enter-label}
\end{figure}


So, what can be done to offset these large wireless signal losses? \
One way to counter this is to somehow have mobile phones and base-stations located closer to each other.
However, this reduces the range of wireless operation and binds the user to a smaller coverage area.
But, this approach can be implemented on a network-wide scale to ensure a large coverage area by performing cell-densification~\cite{bjornson2015designing, gupta2022multiple, auer2011much, austrin2015dense, bjornson2016deploying}.
This entails substituting a single large-area coverage base-station to be replaced by multiple small-area coverage base-stations.
Since the smaller base-stations have to cover lesser area, they also need to transmit at much lower power levels to facilitate the wireless connection.
Further, since the wireless channel is equivalent, the smartphone also can transmit at a lower power-level as well as communicate to the nearby base-station.
Although each base-station on its own transmits at a lower-power, since there are multiple of them, if the number of smaller base-stations grows too large, practical issues arise, such as deployment costs/backhaul, as well the total power consumption ends up overshooting because of a multiplicative increase.
So far, the past-work consensus is that there are perhaps 100s of these smaller base-stations needed to replace a single high-power base-station, and this has led to very limited real world deployments.


In this paper, we present \name, which builds extremely energy efficient connectivity for any given geographical region, and enables greener energy-footprint base-station and longer battery operation of smartphone. \name utilizes explicit environmental features generated via a ray-tracing digital-twin framework (Sionna \cite{hoydis2023sionna}) of the geographical region into account, to strategically place multiple low-power low-coverage base-stations, which cumulatively cover the given larger area at a much lower energy-footprint.
Further, often, acquiring these strategic sites at higher height is challenging, requiring a tower placement, and to address this, \name designs the network with base-stations placed only at 15m height above ground, that can be enabled by mounting base-stations above street-poles or trees. 
In addition to reducing the energy footprint at the base-station, \name's designed network also reduces the transmit power requirements at the smartphone clients, which leads to an extended battery life. We also validate \name's simulation framework in real-world environment via hardware measurements, and further, procure power measurements from various components of commercial smartphone to quantify the exact battery savings.

\name's first contribution is to show that by utilizing the explicit knowledge of the environment, base-station densification can be done in a strategic manner that reduces the number of required smaller base-stations considerably.
To show this, we utilize an open-source ray-tracing framework (Sionna~\cite{hoydis2023sionna}) which can model the wireless environment accurately and can predict signal coverage by simulating EM phenomenon like diffraction, scattering and multipath.
We frame an optimization problem of covering area equivalent of a large power base-station via multiple smaller power base-stations placed in the Sionna's digital environment.
We show that this problem becomes np-hard, and has an analogy to the traditional sum of subsets problem.
Hence, we present two heuristics, a greedy approach, and an hill climbing approach, which perform similar to each other which suggests that the greedy approach could be close to optimum.
Both these approaches do 3-4x better than the previously considered environment-agnostic densification strategy ~\cite{richter2009energy, filo2020performance, kibilda2015modelling, perdomo2020user, ashraf2010improving, hasan2011green}.
Also, the designed approaches are able to minimize resulting interference by minimizing the overlap areas of the multiple base-stations.
We also evaluate the total power consumption (300W) of this optimized network, that comes to be around 700W lower powered than compared to a single base-station consuming about 1000W.
We also show dependence of the densification strategy on height of base-stations and operation frequency, across two different scenes with varying building densities.
We also verify consistency of the Sionna framework via real-world measurements.

\name's second contribution is to extend the designed framework to the client side (smartphones), where we are able to populate the designed small-cell network with smartphones and utilize the path-loss calculations to calculate the resultant transmit powers at the smartphone side.
The simulations reveal a drastic 10-15 dB reduction in the transmit power of the smartphones, since the smaller base-stations are located, on an average 5 times closer to the mobiles.
However, the challenge here is to map this impressive reduction in transmit power to actual savings in mobile battery.
Mobile power measurement campaigns in the past~\cite{xu2020understanding} relied on measuring the cellular power consumption in an indirect manner, by computing power consumption of a task when mobile is in airplane mode and then normal mode, and taking the difference to estimate cellular power consumption.
Although this yields a measurement of average power consumption, it doesn't provide fine-grained measurements.
To this end, we utilize a new tool released by Google to directly procure the On-Device Power Measurements (ODPM) via hardware power rails connected to different modules (display/cellular/compute etc). 
Through a combination of ODPM power rails measurement, and a spectrum analyzer antenna placed right next to mobile for measuring transmit power, we collect combined transmit power and power measurement data from a Pixel 7a smartphone connected to an actual base-station.
This is the first such publicly available dataset with such fine grained hardware measurements, since all the earlier collected 5G performance data ~\cite{xu2020understanding, liu2023close, rochman2023comprehensive,narayanan2021variegated} had primarily focused on core network, or did not collect the transmit power measurements in conjunction with hardware measurement.

In summary, we present a new approach to optimizing locations of multiple small-power base-station which cumulatively cover a single large-power base-station's coverage area, that minimizes the required number of such base-stations from about 100 to about 25-30.
This optimized small-cell network is achieved by placing them strategically and relying on the explicit environment knowledge about the setting given forth by a ray-tracing computational framework.
Further, the achieved small-cell network has a total power consumption of about 300W, that is 700W less than the 1000W single base-station.
In addition, we also explore the benefits which the smartphone clients enjoy because of this small-cell network, which results in 10-15 dB lower transmit power, and about $50$\% longer battery life.





%% file: 2-background_ue_v1.tex
\section{Background \& Motivation}\label{sec:bgm}
\begin{figure}
    \centering
    \includegraphics[width=0.47\textwidth]{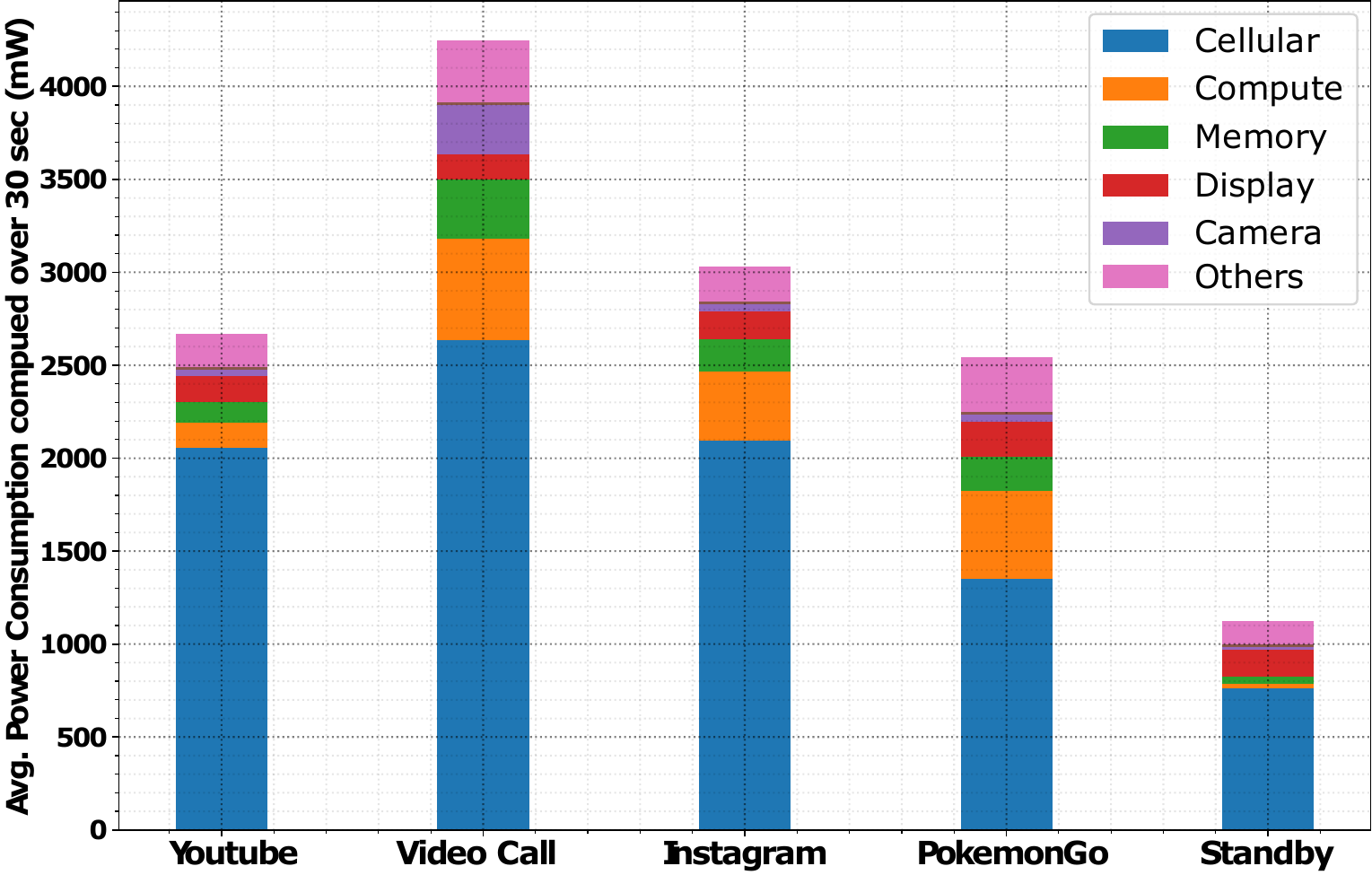}
    \caption{Average power consumption computed over 30-second uses for different activities, like watching a 1080p video on youtube, doing a video call via google meet, scrolling on instagram feed, playing pokemon-go and keeping the phone idle in standby}
    \label{fig:barplot_energy}
\end{figure}

\begin{figure}[t]
    \centering
    \begin{subfigure}[t]{0.5\textwidth}
        \centering
        \includegraphics[width=0.8\linewidth]{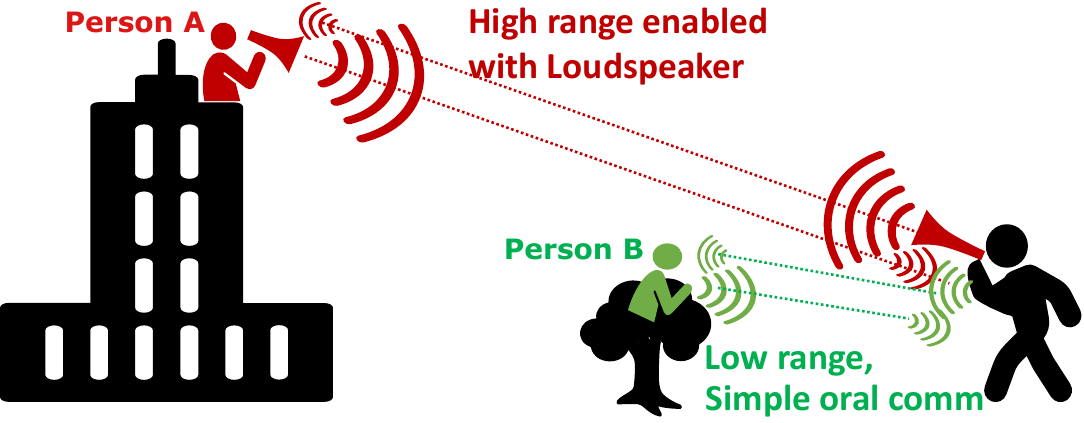}
        \caption{Why communicating far-away incurs more power}
        \label{fig:bgm1a}
    \end{subfigure}%
    \\
    \begin{subfigure}[t]{0.5\textwidth}
        \centering
        \includegraphics[width=0.6\linewidth]{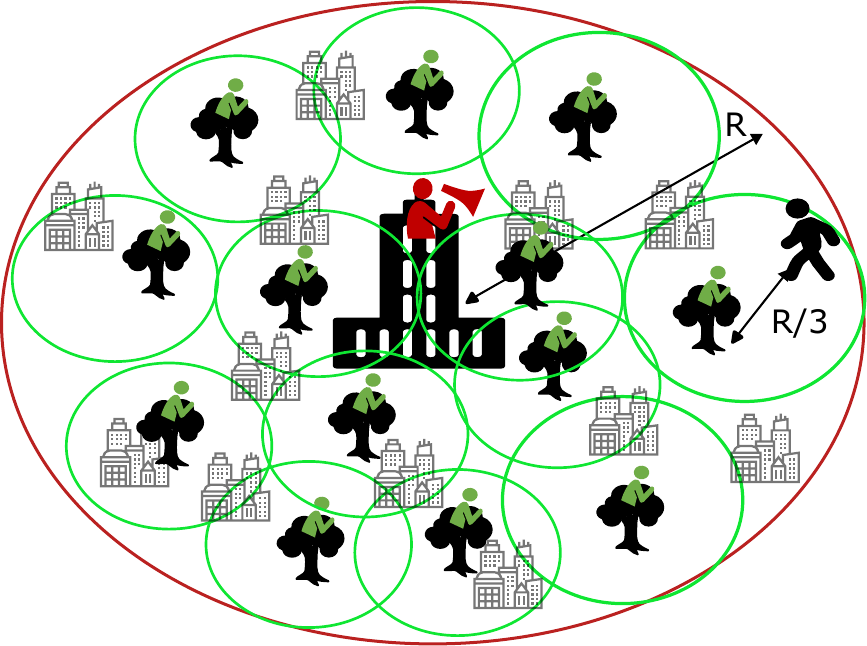}
        \caption{How many Person B's do you need to cover the entire area of Person A?}
        \label{fig:bgm1b}
    \end{subfigure}
    \vspace{1pt}
    \caption{Analogy of densification to simple human oral communication}
\end{figure}

In this section, we will investigate the energy consumption of the user-devices and the breakdown of this energy consumption required for cellular wireless networking. In addition, to that we would also explore as why base-station densification helps with both, improving battery life of smartphones, as well as reducing the total base-station energy footprint.

\subsection{Smartphone power consumption and distance to base-stations}

To motivate why cellular network power consumption has a big impact on smartphone battery consumption, we collect power measurements from different components (cellular, compute, memory etc), while performing everyday tasks like playing youtube videos, scrolling instagram feed and video calls.
Also, we collect data when the mobile is kept on standby, that is, we do not actively use the mobile and it is just turned on and kept idle.
We see that in Fig. \ref{fig:barplot_energy}, cellular power out-trumps all the other components in the smartphone in terms of power consumption.
So, why is it the case that cellular power consumption is so high?

To understand this, consider a toy example in Fig. \ref{fig:bgm1a}, with a person A (red), who is placed on top of a far away building, and is talking via a loudspeaker to you (downlink). Now it's your turn to talk, and in order to reach person A (uplink), you would end up needing a loudspeaker as well, unless the person A has superhuman listening skills (which would also then pick up faint signals from everywhere else).
Compare this to a person B (green), who's just perched on top a nearby tree, and hence both you, and the person B can communicate without loudspeakers. In this analogue, the person A represents a typical macro base-station, which has about 47 dBm transmit power and is typically mounted on a tower. The person B represents a smaller  pico/femto base-station, which has about 17-20 dBm transmit power, and is typically mounted on street-poles/trees or sometimes atop balconies, or rooftops. Even though macro and smaller base-stations have large asymmetricity in transmit power (30 dB, or 1000x lower), their receiver sensitivities are almost the same, or only marginally higher (Macro cell RX sensitivity $\approx -(95-100)$ dBm~\cite{Baicell_outdoor_macro}, Small cells sensitivities $\approx -(92-95)$ dBm ~\cite{Baicell_indoor}), i.e. no superhuman hearing involved. Because of this huge asymmetricity, when a smartphone talks to a far-away located macro cell it needs to transmit at $10-15$ dB higher powers than transmitting to a nearby base-station. 
Since typically, existing deployments utilize a high-power single base-station that is located farther away, most of the population ends up expending high transmit power from their smartphones to maintain the connection, leading to a very high power draw for cellular wireless networking.
In \name, we show how large cellular power consumption can be remedied by having more of smaller power base-stations located close to phones.

\subsection{Why Densification reduces total base-station network power consumption?}

In addition to improving smartphone battery life, we also explain why a network of small power base-stations reduce energy footprint of base-station network. In a typical urban setting, wireless communication between a far-off macro base station and a mobile occurs via a complicated, typically non-line of sight (nLOS) path, because of factors like buildings, elevation differences, and natural landscapes, instead of a direct line-of-sight (LOS) path.
As a consequence, the transmitted signals accrue huge losses due to reflections and penetration through blockages like buildings.
Imagine the same toy-example as before, where you want to hear signals from your far-off Person A, only that you actually can't even see him and can hear faint audio signals of his bouncing off from nearby buildings and reaching you (Fig. \ref{fig:bgm1b}).
Mathematically, this effect has been studied in form of path-loss exponents, which can help understand how the wireless signals decay with distance.
If the the base-station/mobile are located $R$ away, the wireless signal decay as $\propto 1/R^{\gamma}$, with $\gamma$ being the path loss exponent.
Typically with LOS, $\gamma = 2$ as proven in EM-theory, whereas in nLOS, $\gamma > 2$ to mathematically represent the higher losses accrued~\cite{sun2016propagation}.

The key-reason why densification helps reduce energy footprint of base-stations is that $\gamma>2$.
To see this, consider an example where the macro base-station is located $R$ away from the mobile in this scenario, and hence, would experience a loss of say, $\propto 1/R^{3}$, with $\gamma=3$ in this area.
Hence, it would need to transmit at a power $\propto R^{3}$ to offset these losses.
Instead, say, we had a smaller base-station with coverage $\frac{R}{3}$, the smaller base-station would need to transmit with $\propto (\frac{R}{3})^{3}$, $27$X smaller power to reach the base-station.
Hence, the \name question becomes, if we can cover the entire $R$ coverage of a single base-station with $\leq27$, $\frac{R}{3}$ base-stations to save on the transmit power.
Going back to the analogy, say you have Person A located at $R$, and Person B located at $\frac{R}{3}$, to save on the loudspeaker power consumption, and still guarantee you can hear signals across the entire $R$ area, you would need to have $\leq27$ Person B's placed strategically (Fig. \ref{fig:bgm1b}). 
If the considered setting becomes close to LOS, $\gamma\to2$, hence, you would need to somehow put $\leq8$ such close-by Person B's to cover the large area.
Thus, the correct number of smaller base-stations to put depends on the considered scene setting (buildings, elevation, landscapes).

Hence, in this section, we have shown the background behind \name, and why densifying helps conquer both the key problems in today's wireless network: mobile battery draw, and base-station energy consumption. There have been papers that also have advocated for this using stochastic geometry~\cite{ding2015performance,bjornson2015designing}, or following larger scale trend-based approximations~\cite{gupta2022multiple}, however, as we explained, the exact quantification of power-savings depends on number of smaller base-station needed, which also depends on the considered scene geometry.
Thus, in the next section, we will delve into details of how \name is able to the explicit geometrical details of a given scene, and optimize the number of smaller base-stations needed and conquer the carbon-footprint problem in nextG communications.

%% file: 3-design_ue_v2.tex
\section{Methodology} \label{sec:design}

We have so far explained how base-station densification can potentially reduce total wireless network power, as well as reduce the transmit power on the mobile side.
In this section, we present how to utilize an open source ray-tracing based simulation tool (Sionna \cite{hoydis2023sionna}), that can import any large city-scale scene (via OpenStreetMap(OSM) \cite{osm} + Blender \cite{Blender}) to create a framework to optimize base-station locations.
The ray-tracing framework brings explicit knowledge of the environment to the table.
We then show how relying on this explicit knowledge rather than densifying agnostic to the environment, helps minimize number of small base-stations needed by putting the base-stations strategically, and hence reduce the consumed power. 
We present two distinct algorithms to utilize this, one with a greedy approach and other an hill climbing randomized optimizer, both working with about similar performance, which alludes to this problem being np-hard with no easily available global optimized solution.
Finally, we present how the simulation environment can then be extended to compute transmit power requirements at the user-equipment side by populating the obtained densified network with user locations.

\subsection{Why successful network densification needs explicit environment knowledge?}
\begin{figure}[t]
    \centering
    \begin{subfigure}[t]{0.5\textwidth}
        \centering
        \includegraphics[width=\linewidth]{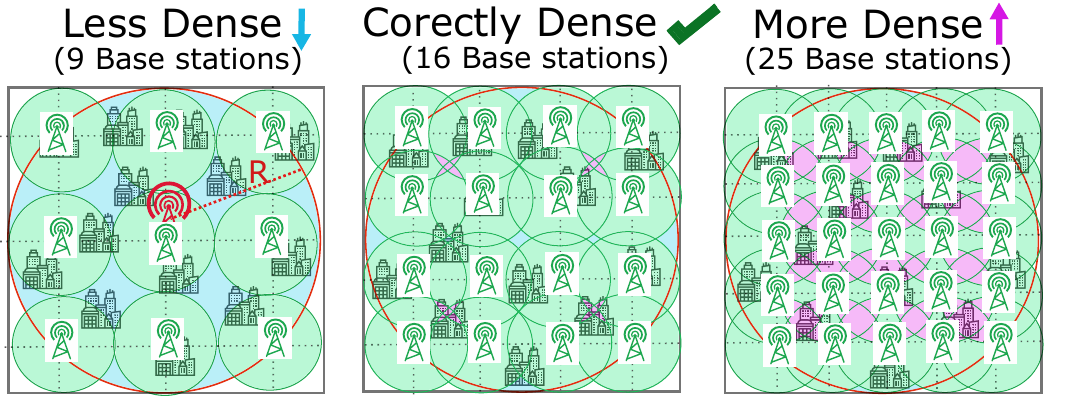}
        \caption{Past work approaches densification problem by performing uniform densification, and increase number of smaller base-stations until there is no blind area (cyan) as well as overlapping areas are reduced (pink)}
        \label{fig:design1a}
    \end{subfigure}%
    \vspace{2pt}
    \begin{subfigure}[t]{0.5\textwidth}
        \centering
        \includegraphics[width=0.9\linewidth]{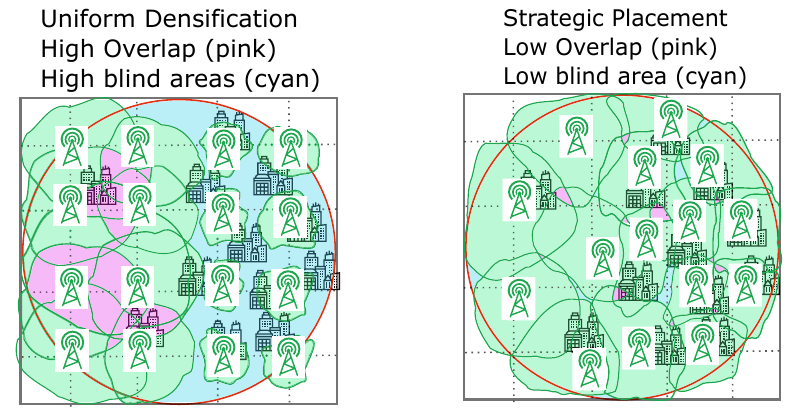}
        \caption{When you take even the best possible uniform densification strategy and apply it to a different urban area, because of inherent asymmetric environment, a strategically placed densification will fare better}
        \label{fig:design1b}
    \end{subfigure}
    \vspace{1pt}
    \caption{Why uniform densification fails in realistic scenarios}
\end{figure}


In this sub-section, we consider two toy examples, to show why simple densification studies based on indirect measures, like path loss trends, are too simplistic to create a working densification strategy for a practical scene, and make a case for requiring explicit knowledge of considered scenario to arrive at an optimized strategy. 

First consider Fig. \ref{fig:design1a}, with building density fairly uniform across the area.
The task at hand is to replace the single macro base-station at the center with multiple smaller base-stations of certain power level. 
Consider we are working with a certain class of smaller base-stations, that only cover a radius of $R/3$, and we need to determine how many such base-stations would be needed to fill an entire $R$ radius area.
The past approaches would basically try to place an increasing number of these smaller base-stations in an uniform way, by dividing the scene into smaller squares, and increasing base-stations from $1,4,9,16\ldots$, and stopping the algorithm when the entire area is covered without large overlaps (pink) and large blind areas (cyan), as detailed in Fig. \ref{fig:design1a}.


However, real world deployments are not uniform, and have natural asymmetry; therefore, now, consider Fig. \ref{fig:design1b} left side, where instead of spread around uniformly, buildings are concentrated near the right area, with left area being more empty. 
If we take the optimized dense uniform deployment now, and apply to this asymmetric setting, base-stations on the left would end up with higher than $2R/3$ ranges, and lead to large overlaps (pink), whereas base-stations on the right will have reduced coverage since building density is higher than expected, and hence, creates large blind areas (cyan).
Instead of densifying uniformly, by knowing the explicit details of the environment, densifying strategically more on the right side and less on the left side, similar to Fig. \ref{fig:design1b} would be much more optimum.
This shows that there is no `one size fits all' approach to densification, and one needs to know the explicit details on the exact urban setting and building distribution in order to arrive to an optimized densification level.


\subsection{Coverage-optimized algorithms, comparison with uniform densification}

\begin{figure}[t]
     \centering
     \includegraphics[width=0.48\textwidth]{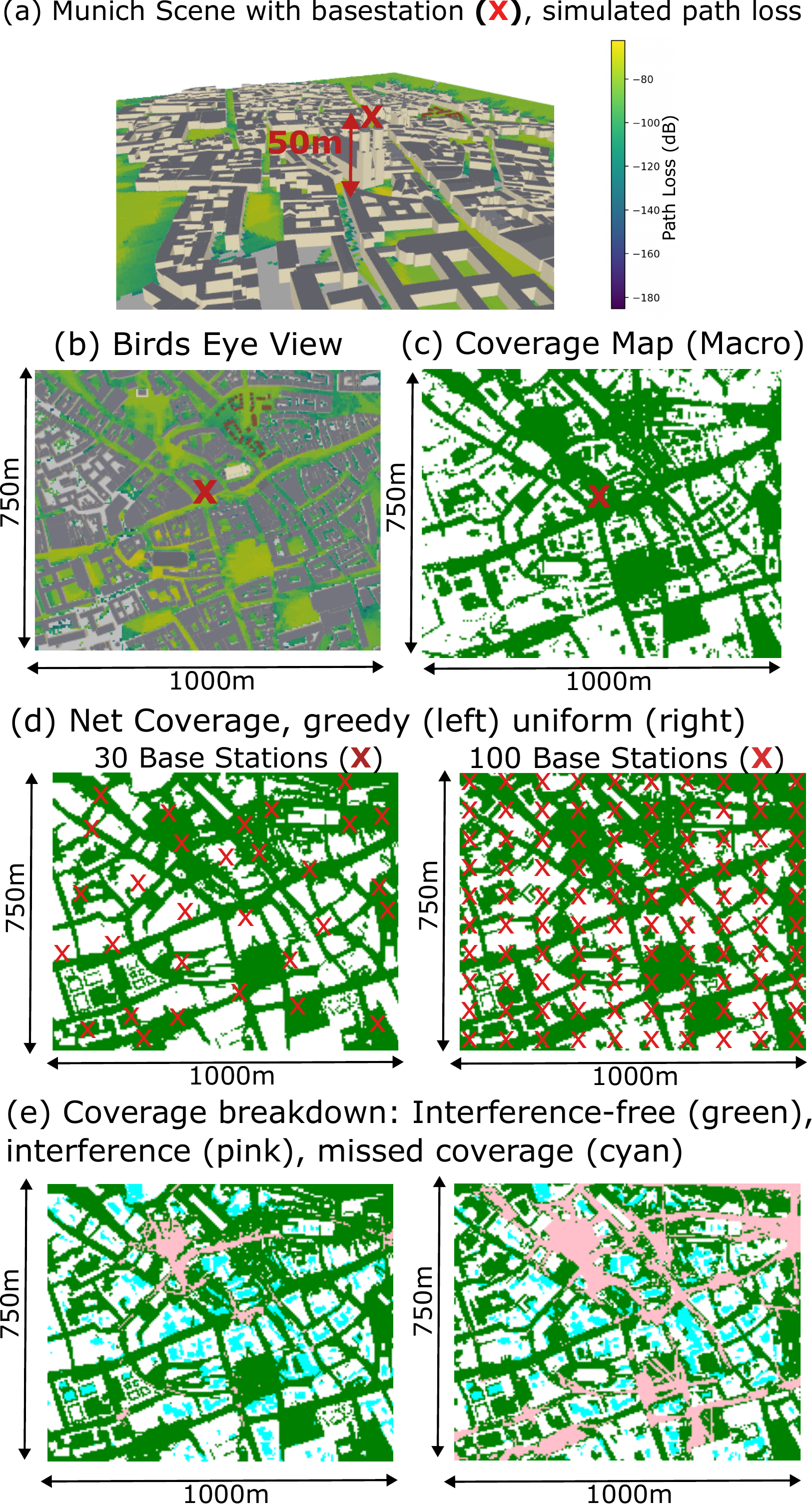}
     \caption{Visual comparison of uniform and greedy algorithms, with (a) describing the Munich scene (b) showing the birds eye view (c) with macro base station coverage map (d) showing cumulative coverage of greedy and uniform algorithms and (e) highlighting the interference and blind areas}
     \label{fig:sionna_scene}
 \end{figure}

So, how can we capture the explicit environmental knowledge of the given scene, and utilize it in a computational framework to arrive at a optimum level of densification? 
The main problem with the past solutions, even including those that take a worse than $2$ path loss exponent into account, is that, the path loss exponent only captures an overall trend of the path loss, obtained via averaging over different paths around a base station and misses the fine-grained explicit details of the environment, crucial for detecting and alleviating blockages and interferences. 
Instead of replying on such indirect metrics, computational ray-tracing tools, in conjunction with fine-grained maps and building data (height, size and shape) can directly compute the coverage itself in a forward pass, and can explicitly showcase blind areas, as well as interference where signal can be recieved from multiple sources.
In fact, today, we have such a conjunction of open source tools available, with Sionna~\cite{hoydis2023sionna} being an open-source ray-tracer that can integrate with Blender \cite{Blender} + OSM \cite{osm}, which allows creating a digital copy of a physical scene and study the wireless channel statistics thereof in the simulations (Fig. \ref{fig:sionna_scene}(a)).

Now, equipped with a ray-tracing tool, how can we approach the optimized densification in a systematic way? Sionna ray-tracing tool can directly measure the coverage map in a forward pass itself, by tracing rays and modelling the electromagnetic phenomenon like diffraction and penetration losses.
In context of densification, it directly allows to compute the coverage of a smaller base-station placed at a certain location, with all the irregularities taken into account.
This makes the method robust to any assumptions, we do not need to compute a coverage radius or simplify the coverage with a circular fit.
By testing multiple locations in the ray-tracing scene, we can directly determine the created scene overlaps, as well as missing areas compared to the single-large base-stations and hence optimize the locations.
But, do we need to test for each and every location to determine the best locations in a brute force manner?

\begin{figure}[t]
     \centering
     \includegraphics[width=0.5\textwidth]{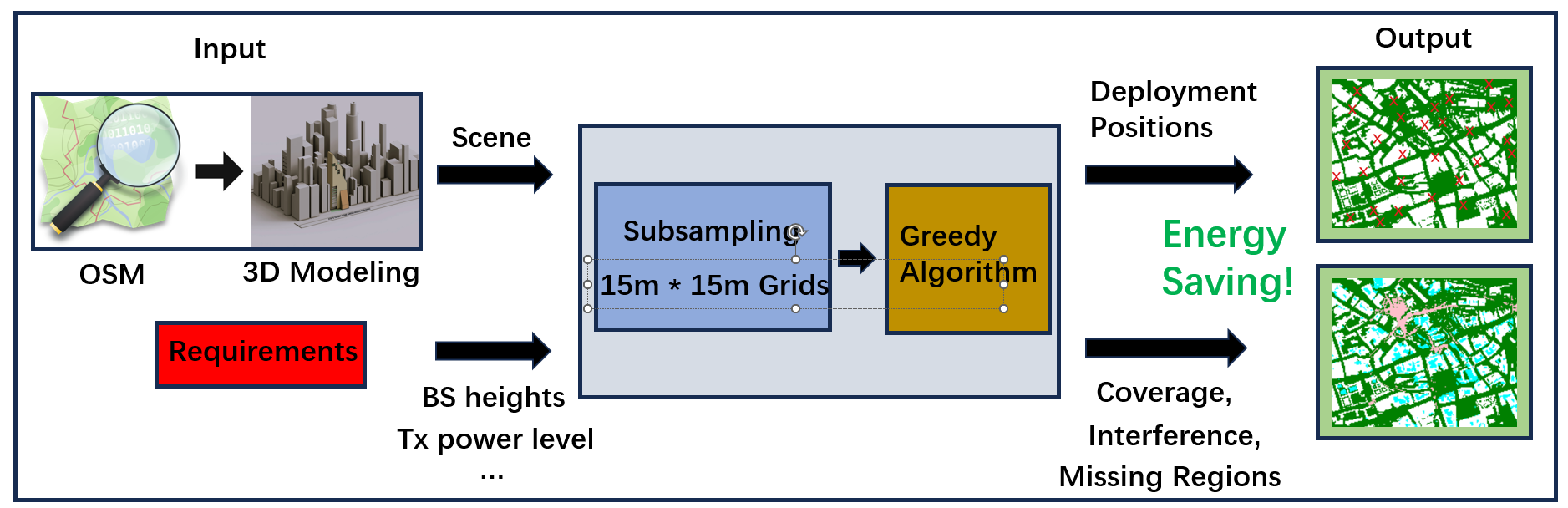}
     \caption{Workflow of the design. \name utilizes Sionna framework that provides building data, takes base-station height and transmit power level as an input to generate a densified network output}
     \label{fig:pipeline}
 \end{figure}

Instead of blowing up the complexity, and performing a brute-force search across all the possible points, we introduce two innovations: First, is a sub-sampling approach which exploits continuous nature of wireless channel and break the entire scene into a discrete grid of $15$m spaced points (that is say we have a $150\times 150$m scene, we will sub-sample to create $10\times10$ grid points), and compute small-cell coverage at each of these points.
Second, once we break the scene into discrete test grid points, we utilize a greedy algorithm to select the $N$ grid points which greedily maximizes the cumulative coverage of $N$ locations.
We keep increasing $N$ until the maximized coverage for $N$ chosen locations reach about $1.1\times$ the coverage area of the single higher-power macro cell deployment.
This way, by utilizing sub-sampling and greedy algorithm, we are able to efficiently compute the minimum number of low-power  low-coverage base-stations, as well as their respective locations, to cumulatively cover the entire large area of a macro base-station.
This process is illustrated in Fig. \ref{fig:pipeline}.


Next, we take a deeper look the the mathematical intricacies behind the greedy algorithm design, why it is close to optimal, and an alternate hill-climbing approach to test the performance of the greedy algorithm. 
It is worth noting that the problem has an analogue to the traditional sum of subsets problem, where the aim is to maximize the sum of subsets beyond a certain given level \cite{alon1987subset, darmann2014subset, austrin2015dense, bansal2018faster}.
However, a twist here is, that the subsets which represent the coverage maps of different chosen locations, may also have intersections, i.e. common areas between them.
Hence, in our greedy step, while going from $N$-th step to $(N+1)$-th step, we minimize the overlap and maximize the new coverage area brought in.
The proposed algorithm to solve the problem regarding the densification is explained in next following steps. The problem itself can be written as follows:
\[min \quad N \]
\[s.t. \quad f(N,\{(x_i,y_i)|i= 1, 2, ..., N\}) < e_m, \]
which N is the total number of low-power base-stations, and \(x_is\) and \(y_is\) are the location of base-stations. In this problem, we consider TX power $17$ dBm, typical of femto cells \cite{Baicell_indoor}, and a fixed height of 15m for all the low-power base stations, which is low-enough to be considered as a street pole or a tree mounting. It should also be noticed that \(f\) is a function whose output is the defined coverage ratio. On the other hand, \(e_m\) is the evaluation metric which can be considered as the ratio of area covered versus total scene area, for a high transmit power macro base station ($47$ dBm, 1000 times more than that of femto) which is located at the height of 50m. In the default Munich scene of Sionna, $e_m = 0.72$, which represents that the macro cell at $50$m covered $72$\% of the entire scene.

\begin{figure}[t]
     \centering
     \includegraphics[width=0.5\textwidth]{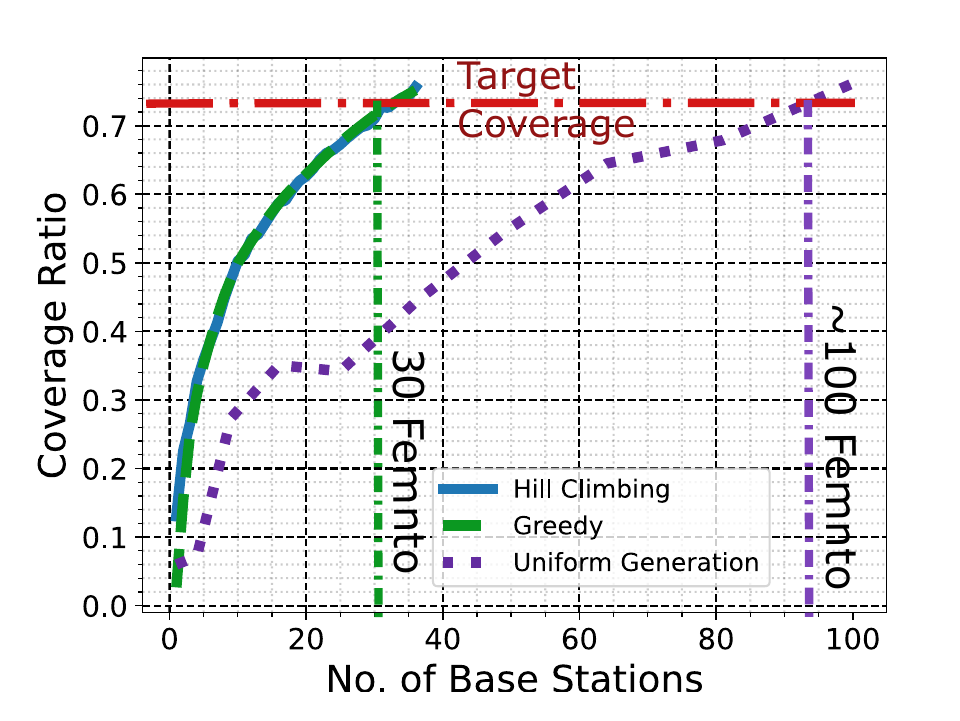}
     \caption{Comparison between different densification algorithms. Clearly, the hill-climbing and greedy algorithms perform similar, and are about 3x better than the uniform algorithm}
     \label{fig:munichDifferentAlg}
 \end{figure}

The greedy step is formulated as follows: the best location for the first base station based on the sub-sampled coverage maps is achieved then for the next base stations, the coverage ratio is just calculated over the area not covered by any previous base stations (subtracting any overlaps). Mathematically, in the \((N+1)\)-th iteration, the formulation will be 

\[max(f(N+1,\{(x_i,y_i)|i= 1,2,..., N\})-f(N,\{(x_i,y_i)|i= 1,2,N\}))\]
\[s.t. \quad f(N+1,\{(x_i,y_i)|i= 1,2,..., k\}) < e_m = 0.72, \]

hence, the algorithm keeps adding base-stations greedily until it covers $72$\% of the scene. Evolution of the algorithm with number of base-stations ($N$) is shown in Fig. \ref{fig:munichDifferentAlg}.
As explained previously, since this problem is analogous to the sum of subset problem, there is no clear global provable optimum solution. To give evidence for the greedy algorithm being optimal, we also try a hill climbing algorithm \cite{russell2010artificial, xi2004smart, jacobson2004analyzing} that selects $N+1$-th stage randomly by perturbing initial solution, instead of greedily. 
That is, for the $(N+1)$-th iteration, $N$ previously added base-station locations are fixed and the $(N+1)$-th location is trying to be tuned in the map. For example, if \(f(j,\{(x_i,y_i)|i= 1,2,..., N\}) > a_{m_{old}}\) which \(a_m\) is the algorithm evaluation metric like what we defined, then \(a_{m_{new}}=f(j,\{(x_i,y_i)|i= 1,2,..., N+1\})\), and for the next iteration \((x_{j_{new}},y_{j_{new}})\xrightarrow{randomly} (x_{j{_{old}}},y_{j{_{old}}})\); Since it explores a random subset of solutions, across multiple iterations, hill-climbing increases the complexity slightly, but we use it as a one-time process to check closeness of greedy solution to optimum solution of hill climbing approach. As seen in Fig. \ref{fig:munichDifferentAlg}, clearly, hill-climbing and greedy approaches perform similar to each other, proving that greedy approach brings us close to the optimum setting.

Further, as compared to the greedy and hill-climbing approaches, which utilize the computational power of Sionna along with sub-sampling approach and optimize coverage directly, we also compare with the past approaches of densifying uniformly.
Clearly, both the coverage-optimized algorithms, which cover $72$\% of the scene, same as that of macro base-station. by placing just around $30$ femto base-stations, outperform the uniform generation method, which requires $100$ base-stations to achieve the same. 
The algorithm outputs for greedy and uniform approach are also shown in the Fig. \ref{fig:sionna_scene}, with the interference and missing areas from macro highlighted in pink and cyan respectively. 

Clearly, one can see that the greedy optimized strategy has lower areas under interference, as well as lower missing areas, while placing $3x$ lower number of base-stations in the process (100 uniform vs 30 strategic).
The reason why there still are blind areas for both algorithms is because when the height is reduced to $15$ m some parts of map remain under blockage, and since Sionna does not consider penetration losses, they remain under blockage even though a base-station is placed next to it. This is a limitation of the framework as of now, however, once Sionna adds penetration, the algorithm description as such would still remain the same.



 \begin{figure}[t]
     \centering
     \includegraphics[width=0.48\textwidth]{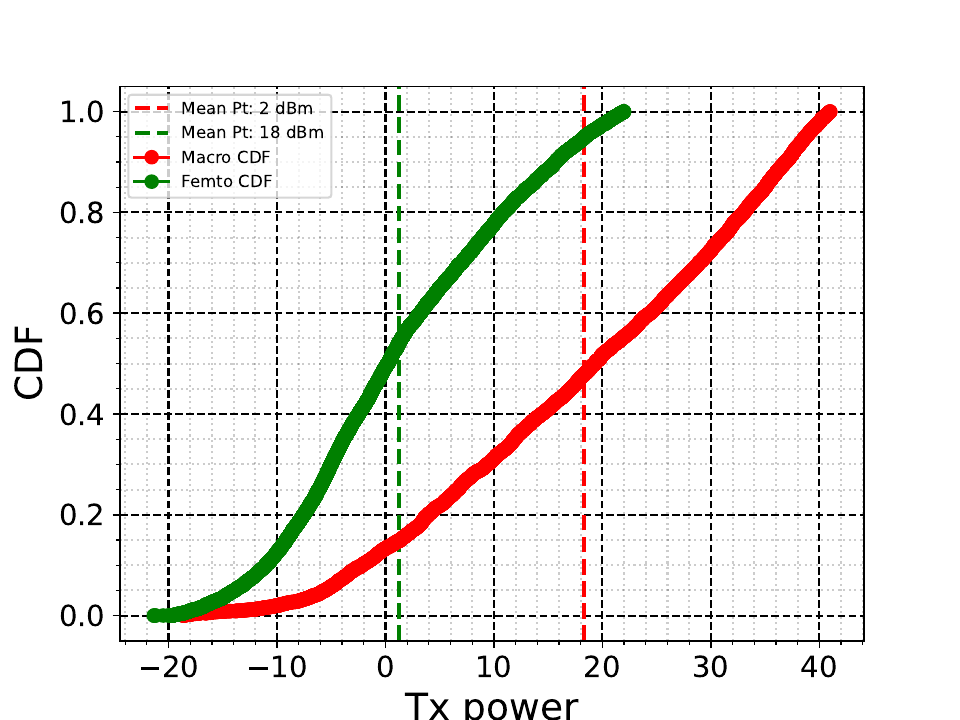}
     \caption{UE transmission power comparison, macro vs 30 femto base stations network, showing about 16 dB reduction in UE TX power}
     \label{fig:munich_3pf_ue}
 \end{figure}

 \begin{figure*}[t]
     \centering
     \includegraphics[width=\textwidth]{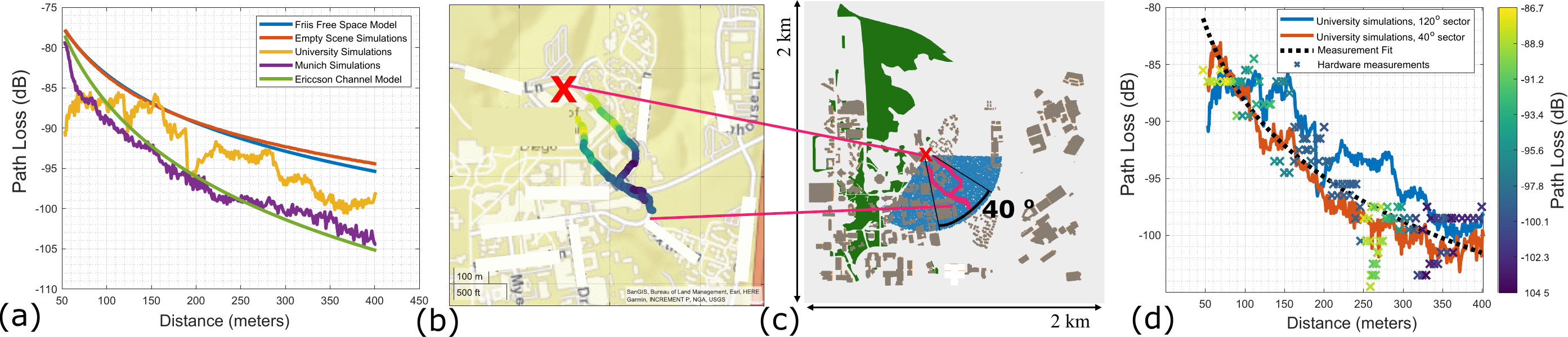}
     \caption{Sionna path loss benchmarking. (a) comparing simulated path loss with Friis Model and Ericcson channel model, (b) university path loss measurement setting, (c) sionna university simulation setup with varying sectors (120$^{o}$ and $40^{o}$) used for improving the computation (d) comparing university simulated path loss with real world measurement, where remarkably, sionna even predicts deep fades around 250m}
     \label{fig:sionna_verif}
 \end{figure*}

\subsection{Extending to user-device simulations and transmit-power requirements}

Moving ahead, we extend our computational network to populate smart-phone devices within the coverage area, and utilizing the path-loss measurements to reverse calculate the transmit power required by the mobile to maintain a given Signal-to-Noise Ratio (SNR) at the base-station.


More specifically, we populate a large number of smartphones ($10000$) randomly scattered across the coverage area for both single macro cell base-station placed at $50$m, and also the $30$ femto-base stations with locations output from the greedy algorithm. 
We calculate the transmit power required by the mobile to offset the path loss, and maintain a SNR of about 15 dB to the base-station which has highest received power at the UE (typically the UE would select the best base-station from which it gets highest receive strength).
The assumed sensitivity of macro base-station is -100 dB (About $3$ dB lower than reported commerical levels \cite{Baicell_outdoor_macro}), and femto base station -90 dB (About $3$ dB higher than reported commercial values \cite{Baicell_outdoor_macro}), since the macro may have slightly better hardware and more sensitive.
However with even 10dB unfairness given to sensitivity, we see that across the $10,000$ randomly generated points, the computed mean transmit power needed for macro deployment is $18$ dBm, whereas for femto is $2$ dBm, showing about a $16$ dBm reduction.
Further, note that this impressive reduction happens even if we are slightly unfair to femto base-stations, since the modern 5G base-stations have similar sensitivities across different class of base-stations \cite{Baicell_indoor,Baicell_outdoor_macro}, however since the gains are impressive, it also entails femtocells can be made with lower sensitivites to save some R\&D cost if need be.

%% file: 5-evaluation.tex
\section{Evaluation}\label{sec:evaluation}


In this section, we first present hardware measurements of path-loss with a base-station deployed in our university, which confirm that the sionna simulation framework by at large follow the hardware measurement trends.
After confirming the sionna framework, we present an estimate on net power savings at both smart-phones and base-stations made possible with \name approach of densify and conquer. 
Specifically, on the smartphone side, we collect logs from an actual smartphone (Pixel 7a) connected to a commercial base-station, as well as characterize its transmit power with a spectrum analyzer, to showcase the possible power savings due to smartphones being located closer to base-stations.
On the base-station side, we present a case-study by looking into various classes of base-stations (macro/micro/pico/femto) and their respective power consumption to translate the results of our ray-tracing simulations to estimated power savings.


 

\subsection{Verfying the path-loss modeling of Sionna}

As detailed in Section \ref{sec:design}, we utilize Sionna ray-tracing computational framework to bring-in the explicit features in a given urban environment. 
The sanity of these simulations rely on correct modelling of the wireless signal path loss with distance, and hence we present results that verify if the simulations follow the trends observed in hardware measurements.

In Fig. \ref{fig:sionna_verif}(a) we show that the path-loss observed in the default Munich scene in Sionna by-and large follow the popular Ericsson channel model~\cite{milanovic2007comparison} used in typical urban scenes. Another important point to note is that when we simulate an empty scene in Sionna, it follow the theoretical Friis path loss equation that theoretically predicts $R^2$ decay.
This proves that sionna is able to adjust to a particular scene and calculate path-loss behaviour for a particular scene.
Proceeding ahead, we collect path-loss and GPS measurements using LTE discovery android app~\cite{lte_discovery} from a base-station deployed by our university dedicated for wireless experimental research in CBRS band (3.5 GHz), by walking uptill 500m in a $40^{o}$ sector, as plotted in Fig. \ref{fig:sionna_verif}(b), with the imported scene in Sionna (via OSM~\cite{osm} and blender~\cite{Blender}) shown in \ref{fig:sionna_verif}(c).
The path loss hardware measurements, fit and sionna simulations are shown in Fig. \ref{fig:sionna_verif}(d).
It is notable that when we zoom into the exact $40^{o}$ sector in the simulations, sionna is able to fit very well with the hardware measurements, and is also able to predict some fading losses observed at 250m mark.
Since, our university setting is more spread out and sparser than Munich scene, and Sionna correctly puts the curve in middle of the empty scene and Munich scene measurement.
Further, we see that the hardware measurements follow the trend captured in simulations as well which further confirms that Sionna is capable in modelling the explicit details of a given map.


     
    
    
    

 \begin{figure*}[t]
     \centering
     \includegraphics[width=\textwidth]{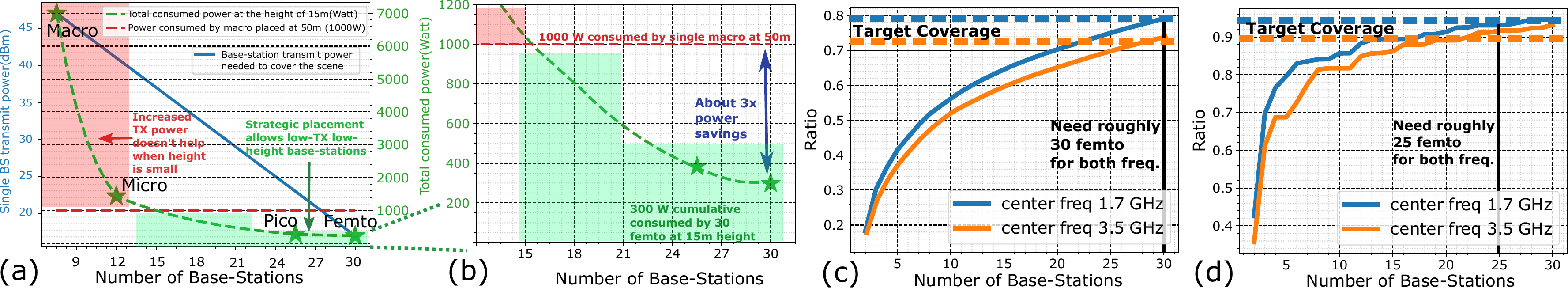}
     \caption{(a-b) Power savings for Base-station side: We first compute the blue curve, that represents how many base-stations are needed to cover the scene for a particular chosen TX power. Then we compute the green curve via 4 discrete points chosen to be the standard macro/micro/pico/femto definitions. The green curve shows cumulative power consumption vs number of base-stations that shows flattening trend at around pico-femto level, where we observe 3x energy savings. (c-d) Show that the number of base-stations and the algorithmic trend remains by and large similar for different frequencies and scenes considered}
     \vspace{2pt}
     \label{fig:combined_bts_pow}
 \end{figure*}
 
 \begin{figure}[t]
     \centering
     \includegraphics[width=0.5\textwidth]{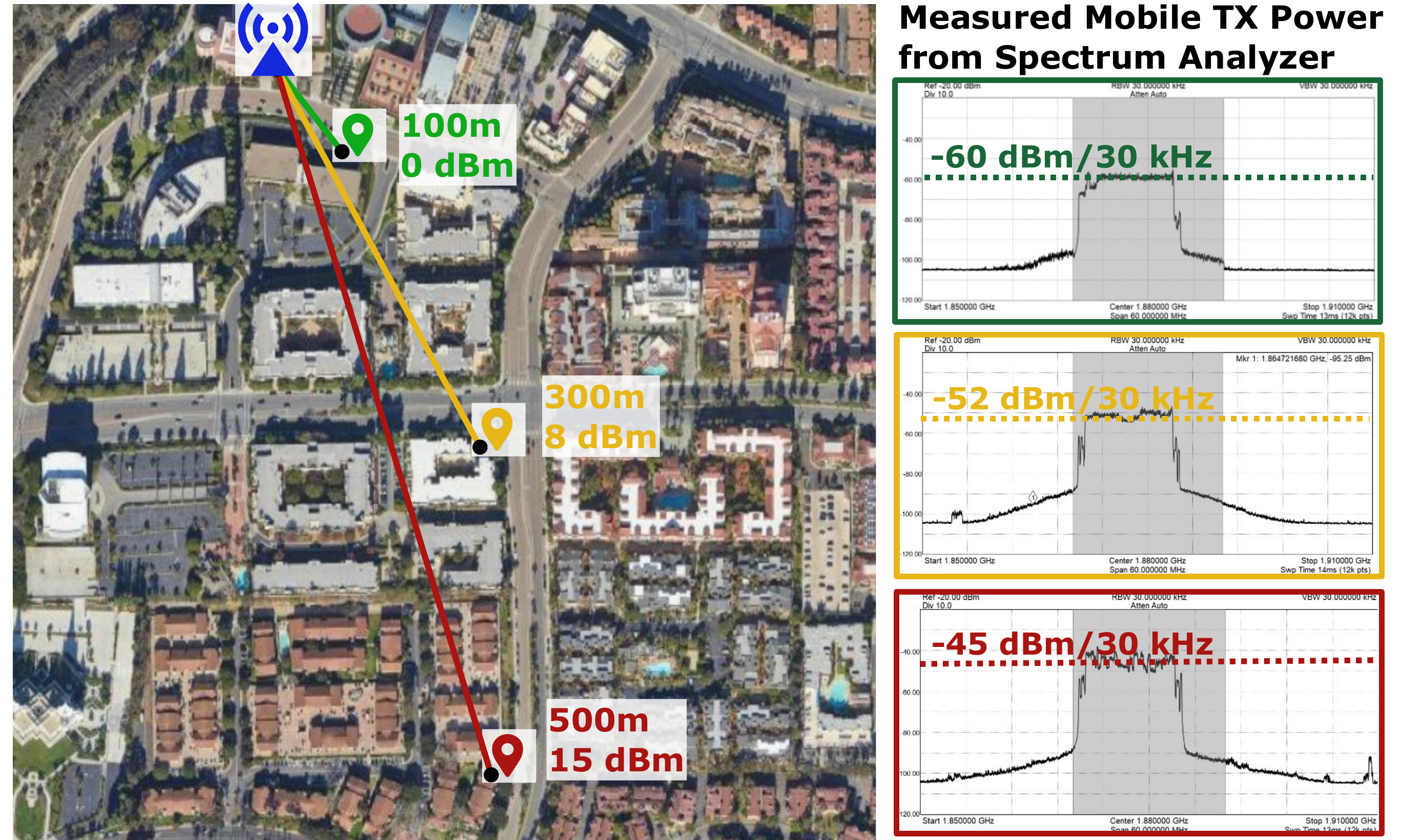}
     \caption{Test locations GPS plot, located 100m, 300m and 500m away respectively, along with the Signalhound measured transmit powers}
     \label{fig:testLocIperf}
 \end{figure}

 \begin{figure*}[t]
     \centering
     \includegraphics[width=\textwidth]{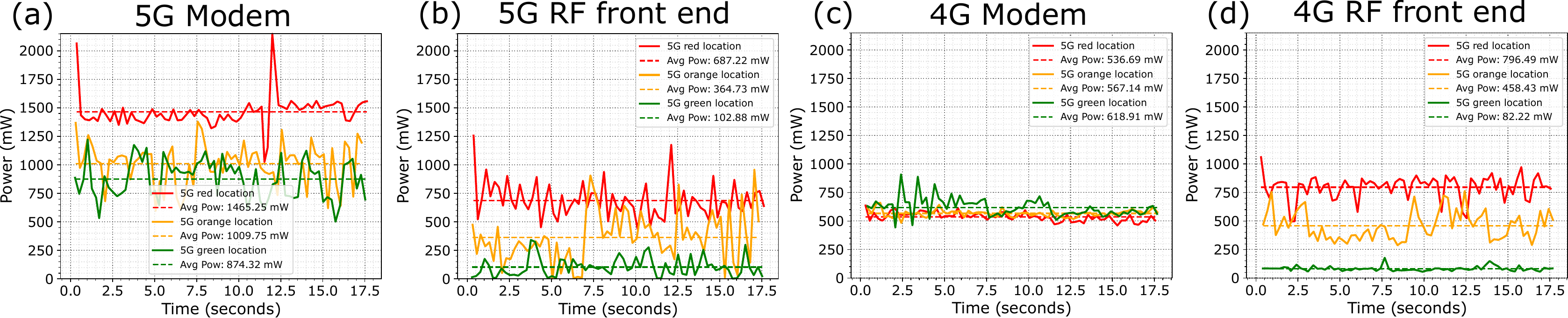}
     \caption{Cellular Power measurements on Pixel 7A at 3 locations shown in Fig. \ref{fig:testLocIperf} for 5G and 4G, (a) 5G Modem power increases about 2x from green to red, (b) 5G RF front end power increases about 7x, (c) 4G Modem power remains about same, decreases slightly due to reduced data rate supported, (d) 4G RF front end power increases similar to 5G as user moves farther}
     \vspace{2pt}
     \label{fig:RFEIperf}
 \end{figure*}

\subsection{Base station power analysis}

In this subsection, we will evaluate the total power consumption for the $30$ femtocell base-station network, all located at $15$m height in the Sionna scene, and also vary the transmit power to find the behaviour of how many base-stations are needed to populate the scene for a given transmit power. 
Then, we utilize popularly reported power measurements for different class of base stations from \cite{auer2011much} to convert the tranmsit power versus number of base-stations curve, to a net power consumption curve, in order to better quantify the energy footprint savings at the UE side.
We also show similar measurements for the university scene, and also two different frequencies, to show that the performed computation generalizes fairly well.

As shown in Fig. \ref{fig:combined_bts_pow}(a), by lowering the height of the base station to $15$ m, covering the target $72$\% of the scene requires placing $7$ Macro base stations, since lowering the height creates blind spots scattered around, requiring an increase and decentralized distribution of the macro base-station. 
Thus, it makes sense to lower the transmit power and then densify, and we see a transition from increased power consumption because of reduce height, versus power savings near micro base-station level.
As we go further to femto and pico, the power consumption flattens out to about $300$ W, which is $3$x lower than the power consumption of a single macro base-station at 50m height (1000W).
The green curve is calculated by taking the known transmit powers of different base-station classes (macro: 47 dBm, 1000W consumption; micro: 38 dBm, 144W consumption; pico: 21 dBm, 14.7 W consumption; femto: 17 dBM, 10.4 W as used in prior work \cite{auer2011much,gupta2022multiple}), and multiplying with number of needed base-station in blue curve of Fig. \ref{fig:combined_bts_pow}(a).
The blue curve basically tells how many base-stations are needed to gurantee $72\%$ coverage area for a given chosen transmit power.
For example, if a base station with 17dBm transmit power should be employed, 30 base stations are needed to achieve the coverage metric satisfaction which needs \(30\times 10.4=312W\), something more than 3 times power saving over the single macro base-station.
However, note that we have not computed power required in backhauling, and as a result the expected power savings might be slightly lower.
It has been shown that wired networks are about $10$ times more energy efficient than wireless networks, and hence our expectation is that a wired backhaul would not add a lot of energy requirements to the $30$ femtocell deployment strategy\cite{baliga2011energy}.

 



Now, when we know the number of base stations to achieve the similar performance as a Macro base station located at the height of 50m, we can see how greedy algorithm by adding base station try to increase the coverage metric beyond the required $72$\% requirement, in an optimum way. As shown in Fig. \ref{fig:combined_bts_pow}(c), by increasing the number of femto base stations, the coverage metric is also increasing till the point enough number of base stations has been added. Till this point, all the simulations were done in center frequency of 3.5 GHz. On the other hand, this simulation can be done for other center frequency such as 1.7GHz, and we can see due to lower path loss in this frequency the ratio curve with adding the similar number of femto base stations will be above the 3.5GHz frequency; however, due to the coverage metric used in greedy algorithm which itself is a function of frequency, the total number of base stations to satisfy this metric wont be very different for two different frequencies. Also, we repeat the simulations at these two frequencies for the self-generated university scene that we previously used for sionna verification. Since this scene is sparser than the Munich scene, the coverage achieved by the macro cell bumps to about $90$\%, and hence the target coverage also increases. The algorithms evolve in a similar manner, and show that in the university scene, the same performance can be met with $25$ base-stations.
This shows that the greedy algorithm presented in
the paper generalizes fairly well to different scenes and frequencies of operation.

\subsection{UE power saving verification}

Next, we present the possible power-savings on the smartphone side as a consequence of connecting to a closeby base-station as compared to a farther away base-station. 
Further, we want to quantify the possible power-savings at the smartphone side and map the predicted 15 dB transmit power savings to impact on battery life.

\begin{figure}[t]
     \centering
     \includegraphics[width=0.5\textwidth]{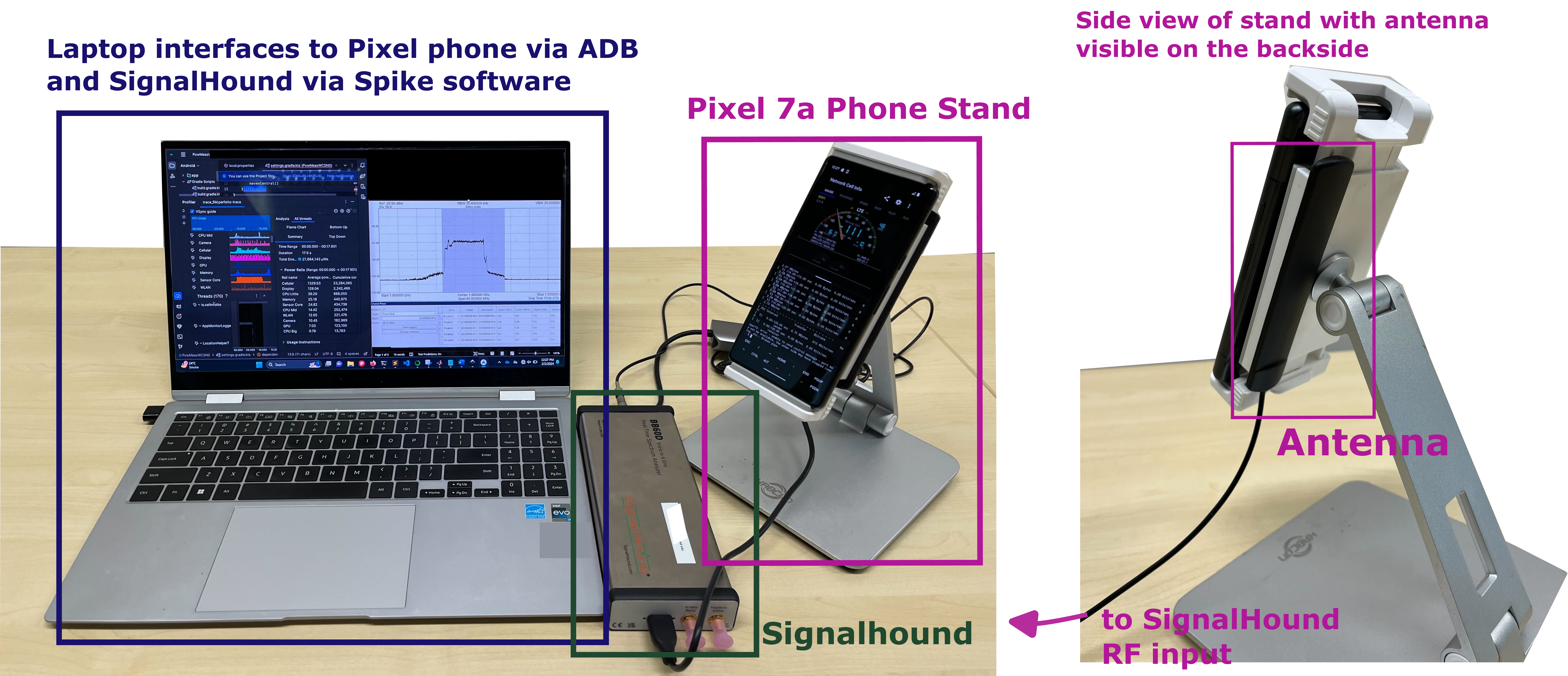}
     \caption{Setup to measure both the transmit power of phone, as well as power consumption reported via ODPM rails}
     \label{fig:setup}
 \end{figure}

Furthermore, by collecting system traces from an actual smartphone connected to a commercial base station, at three locations (100m 300m and 500m, GPS plot shown in Fig. \ref{fig:testLocIperf}) as well as measuring the mobile transmit power, we quantify how much savings are possible because of densification.
To enable this, we create a measurement setup consisting of (1) Pixel7A phone~\cite{pixel_7a}, (2) Signalhound BB60D spectrum analyzer~\cite{signal_hound} connected to an antenna mounted next to Pixel7A and (3) Windows Laptop connected to both the components and pulling in the system logs from android phone via ADB and spectrum analysis data from the signalhound. 
The setup image is shown in Fig. \ref{fig:setup}.
Pixel phones 6 and above~\cite{Power_profiler} implement the On-Device Power Measurement (OPDM) power measurement rails, which report the hardware measured power of different components (Cellular consisting of RF sub-6 GHz front end, modem, mmWave front end, display power, compute power consisting of CPU+GPU, memory access power etc).
To ensure that we collect cellular power data, we initiate a 5Mbps iPerf3 via Termux shell~\cite{termux} on the phone, let it stabilize and then initiate a system trace collection.
We utilize Perfetto python API~\cite{perfetto} to parse through the system trace and obtain the power rails measurement reported by OPDM.
In conjunction with transmit power measurements via the spectrum analyzer, we are able to see a reduction in front end power (which synthesizes the transmit signal) by about 7 times, as transmit power is lowered by 15 dB (Fig. \ref{fig:RFEIperf}(b)).
Further, the modem power reduces by about 1.5 times (Fig. \ref{fig:RFEIperf}(a)), which could be because of MIMO modes getting enabled as the mobile goes farther. 
In these measurements the only thing which changes is the distance between the smartphone and base-station, and hence this increase in power consumption is clearly because the modem+RF front end trying to offset the larger path losses, which is then also confirmed by our spectrum analysis via the signalhound.

\begin{figure}[t]
     \centering
     \includegraphics[width=0.48\textwidth]{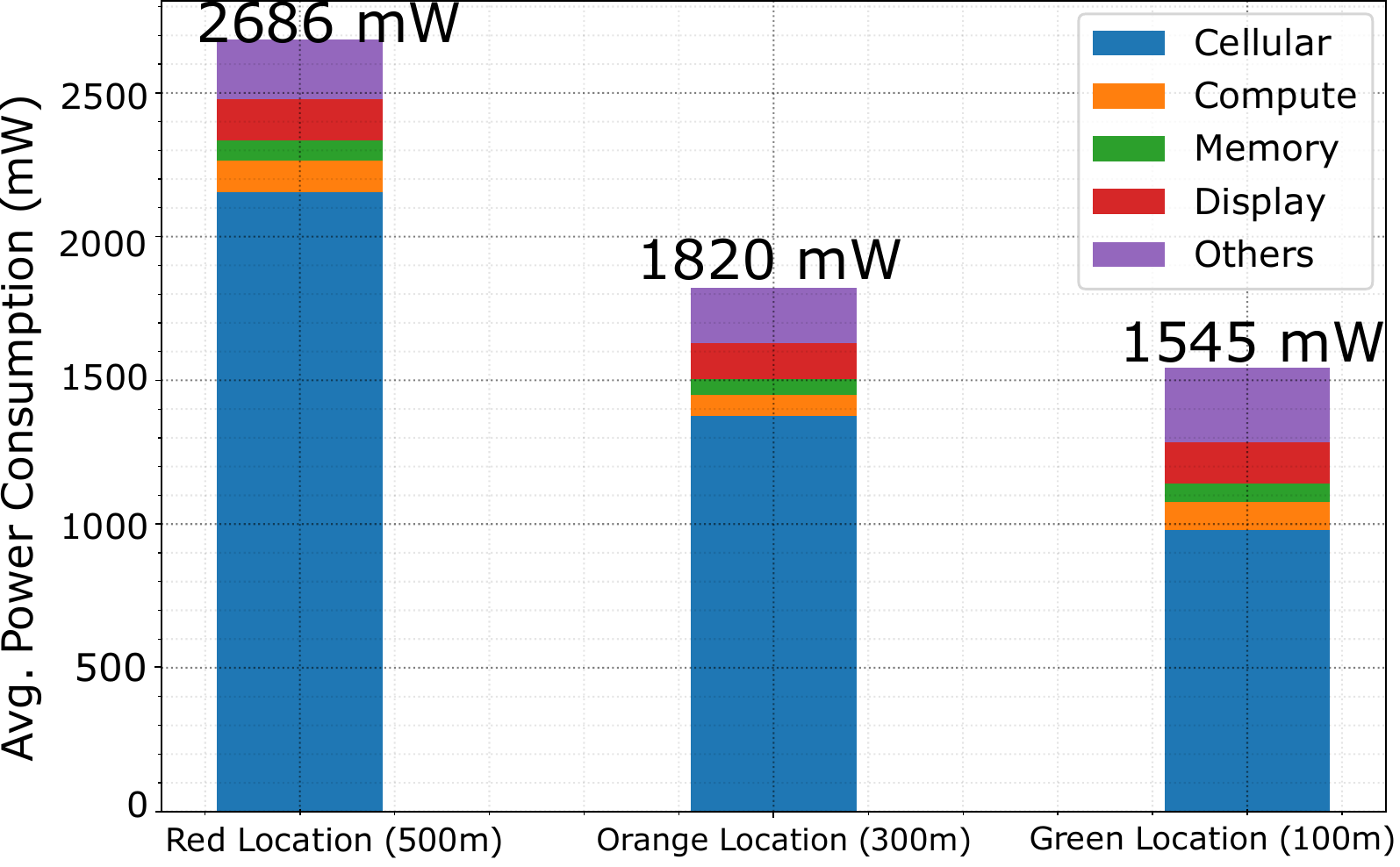}
     \caption{Average power consumption while doing iPerf for 5Mbps at the chosen three locations}
     \label{fig:bar_plot}
 \end{figure}

We also collect similar results for 4G, where the measurement shows similar reduction in RF front end power as the base-station (Fig. \ref{fig:RFEIperf}(d)), however in 4G there are two differences (1) in 4G at medium/far locations, the PHY layer is unable to deliver 5 Mbps, and the modem is only able to support 2/0.5 Mbps respectively (2) as a consequence, modem power decreases slightly (Fig. \ref{fig:RFEIperf}(c)), whereas in order to offset the larger path losses, the RF front end power shows similar trend to 5G.
Hence, it is clear from these measurements that 5G is able to support the 5Mbps throughput everywhere, however at an expense of increased power-consumption.
We also show the total average power consumption across all the different power rails in Fig. \ref{fig:bar_plot}, computed over 30seconds, and clearly, the cellular power consumption is the highest contributor in 5G phones.
With our measurements, the 5G phone while communicating 5 Mbps data, consumed on an average 2600mW when it is located 500m away, and 1500mW when it is located 100m away.
Hence, considering the 5800mAh battery capacity (3.7 V nominal voltage, which makes it $\approx$21Wh), the 5G phone located 500m away will expend the total battery in 6 hours, whereas it will take 9 hours if it is located 100m away.
Typical range of operation for femtocell is 100m, and thus, in our proposed femtocell network, the maximum range will be around 100m.
This implies that i the femtocell network, the smartphones will observe about a 50\% increase in the battery life when they communicate via the cellular network, and about 7-8 times reduction in the RF front end power consumption, as well about 1.5 times reduction in the modem consumption.

%% file: 6-limitations.tex
\section{Discussion and Limitations} 
\noindent \textbf{Sionna Inaccuracies in very far distances (>1km)}: 
In the paper, we show that Sionna is able to map the wireless channel path loss really well for distances $<500$m. However, for farther distances ($>1$km), the generated rays by sionna become sparser and the framework predicts inconsistent results. 
In this paper, we utilized an in-house server with $4$ A10 Nvidia GPUs to run sionna.
However, to gurantee dense rays at $1$-km distance, the number of rays required would be 30e6, which would expend the memory limit of an Nvidia A10 GPU.
however with a larger scale compute, simulating more than $1$ km should also be feasible. 




\noindent \textbf{Indoor, Elevation not considered}:
In the design, we optimize the base station deployment only considering the regions which are outside the building. 
This assumption is largely fine, since cellular connections are typically meant for outdoor coverage.
However, there is a trend towards indoor 5G, and in the future, we can attempt augmenting Sionna dataset with indoor hardware measurements.
Further, right now Sionna has only flat ground plane, whereas in reality landscape would also play a major role in base-station location selection.
This requires better environment modelling, and perhaps some hills can be modelled as smoother buildings, and valleys can be modelled as an area surrounded by higher buildings.
However this is left as a future work for now.

\noindent \textbf{Adding MIMO to the framework}:
Refer to Fig. \ref{fig:sionna_scene}, there is still undeniable interference happening in middle (pink spot), even though our design minimizes the interference. However this could be solved with MIMO by enabling spatial isolated beams supporting users in pink spot. Further, the algorithm can be augmented to work with MIMO which means adding one more variable to the optimization output that is the beam pattern of the base station, which can further help with adding more nLoS path to the coverage area of smaller cells and increase the coverage of the base stations as well as reducing the interference of the base stations. 

\noindent \textbf{Unconfirmed MIMO measurement results}:
Refer to Fig. \ref{fig:sionna_scene}(e) we did UE side power measurement comparing 4G/5G on Modem and RF front end. Although we can explain that in both 4G/5G cases, due to distance, the further UEs need to transmit more power (as shown RF front end power consumption increases as distance increases), however the significant increasing in Modem power consumption, which happens only in 5G case, could stem from higher layers of MIMO being enabled\cite{gupta2023greenmo}. However, this needs to be verified by further studies.

\noindent \textbf{Downlink cellular power consumption}:
Although in this work, we showcase importance of considering cellular power consumption (Figs. \ref{fig:barplot_energy}, \ref{fig:bar_plot}), the benefits of densification are mainly limited to saving uplink power consumption. There is however, an extra aspect of MAC layer improvement that needs to be explored further. It could be that if smartphones are located closer to the base-stations MAC can become more efficient and lead to power savings. However, this requires finer data collection and improved setup that can provide MAC layer logs, like Keysight WaveJudge \cite{wavejudge}.

\noindent \textbf{Backhauling requirements}:
It is undeniable that success of the densified network would depend on a good backhaul network, and this can be either a wired backhaul by utilizing the already vastly laid, and underutilized optical fiber network \cite{10.1145/3596262}, or try to do a wireless backhaul using upcoming technologies like mmWave fixed wireless access (FWA) \cite{aldubaikhy2020mmwave,de2023mmwave}. A rigorous analysis is needed by adding bakchaul power consumption to better quantify the power savings at base-station network level by considering backhaul network as well.


%% file: 7-related.tex
\section{Related Work}
Path loss, as a key motivation of our work, has been also studied by other researchers from different perspectives. In \cite{li2023geo2sigmap}, authors have tried to enhance the simulated path loss result by combining machine learning and real world data. Also in \cite{ruah2023calibrating}, authors have tried to calibrate the material properties using local phase error estimates. 

Due to its promising advantages (like energy efficiency, throughput, etc), densification for base station deployment is always a hot topic in nextG communication. In \cite{gupta2022multiple}, authors defined PLE parameter as a metric which can be used to figure out to what extent cell densification can work toward decreasing power consumption by base stations as well as showing the simulation result which confirms the power consumption drop. Moreover, authors in \cite{ge20165g} and \cite{richter2010cellular} analyzed the exetent of cell densification which can improve saving the power. On the other hand, in some other literature, results showing the benefits of using cell densification exist. In \cite{richter2009energy}, authors used area power consumption as a system performance metric to assess the performance improvement of using micro base stations replacing macro base stations. Moreover, authors in \cite{ding2015performance} and \cite{jafari2016performance} showed the effect of LoS and nLoS impact in dandified cellular network. 

Researchers also tried to understand the UE side power savings. As discussed in \cite{hoadley2012enabling}, user distance from the base station can decrease the maximum bit rate as signal strength gets lower and noise increases which enforce using high signal power. Unfortunately, users usually are located near edges which can make this evidence closer to  reality. Although this paper talks about the improvements can be achieved using cell densification, no solid result using this approach is mentioned. Moreover, authors in \cite{hoydis2011green} showed the potential benefits the using small-cell networks can bring into system, however, they also take into account the challenges such as interference management and mobility.



Furthermore, authors in \cite{ramanath2009optimizing} characterized the throughput as a function of the cell size and showed that densification which can lead to a higher throughput is also dependant on other factors. However, this work did not consider anything beyond free space path loss such as shadowing and fading. In \cite{park2014asymptotic}, authors showed the spectral efficiency is a logarithmic function of base-station density as number of base-stations grows.


As well as power saving capability at the base station side, some papers researched about tecniques which can lower the power consumption. In \cite{yu2016towards}, they propose a sleep/awake mode technique which can increase the power saving in base station side. Furthermore, in some works like \cite{haider2019maximum}, authors showed the impact of the cell densification on UE side. However, they just considered three different scenarios. 

Other than our approach in saving base station power, researchers also believe that techniques like sleep-mode control \cite{feng2017base, wu2015energy}, MIMO \cite{lu2014overview, prasad2017energy} can also make base stations energy efficient.. 



%% file: 8-conclusion.tex
\section{Conclusion}

In this paper, we presented \name, that utilizes a ray-tracing framework to strategically place densified, low-power base-stations that cumulatively cover the entire area of a high power base-station.
We show that such a densification strategy conquers two key problems, it can potentially save about 3x power savings by avoiding large wireless signal losses due to high range, and also lead to $50$\% improved battery life for the connected smartphones on the dense network as a consequence of being located closer and at a smaller height to them.
\name is able to quantify these impressive gains because it is the first work performing base-station densification studies by utilizing the recently released Google's OPDM power rail measurement tool to profile hardware power measurements at android phones, and NVidia Sionna open source ray-tracing framework.
With increasing carbon footprint of wireless networks, and primary contributors being the operating energy costs of base-station and the embodied footprint of smartphones, we believe \name approach can lead to a more greener development of wireless networks, as we leap forward to 6G in the upcoming years.


%% file: 9-acknowledge.tex
\section*{Acknowledgment}

This work was supported in part by the U.S. National Science Foundation (NSF) under Award No. 2030245. Further, the authors acknowledge the WCSNG group members for their feedback.